\documentstyle[11pt,newpasp,twoside]{article}
\def\edcomment#1{\iffalse\marginpar{\raggedright\sl#1\/}\else\relax\fi}
\reversemarginpar

\begin{document}
\title{Baryon Oscillations in the Large Scale Structure}
 \author{Asantha Cooray}
\affil{California Institute of Technology,
Pasadena, California 91125, USA.}

\begin{abstract}
We study the possibility for an observational detection  of 
oscillations due to baryons in the matter power spectrum and
suggest a new cosmological test using the angular power spectrum of halos.
The {\it standard  rulers} of the proposed test 
involve overall shape of the matter power spectrum and 
baryon oscillation peaks in projection, as a function of redshift.
Since oscillations are erased at non-linear scales, 
traces at redshifts greater than 1 are generally preferred. 
Given the decrease in number density of clusters at high redshift, however,
one is forced to use tracers corresponding to galaxy groups and 
galaxies  themselves.
\end{abstract}

\section{Introduction}

There is now increasing evidence for 
the presence of oscillations in the angular power spectrum of
cosmic microwave background (CMB) anisotropies. 
The matter power spectrum of the large scale structure 
also contains a signature of baryons in the form of oscillations;
the amplitudes of oscillations, however, are significantly lower with 
fluctuations at the level of $\sim$ 5\% for $\Lambda$CDM
with widths of order $\Delta k \sim 0.02$ h Mpc$^{-1}$. For an unambiguous
detection, 3d surveys should have resolution scales greater than
$L \sim 2\pi/\Delta k \sim$ 300 h$^{-1}$ Mpc in all three dimensions.
In 2d, oscillations can also be detected via the projected power spectrum.
The 2d power spectrum requires no precise redshifts and can be constructed with
halos detected in upcoming wide-field lensing and SZ surveys.

Following our suggestions that the angular power spectrum of halos 
can be used as a probe of cosmology (Cooray et al. 2001), 
projected oscillations provide a new cosmological test; this is 
similar to the measurement of the angular diameter distance to the 
last scattering surface using the first acoustic peak 
in the CMB anisotropy power spectrum.

\section{Halo Clustering as a Probe of Distance}

The angular power spectrum of halos in a redshift bin is a 
projection of the the halo number density power spectrum 
\begin{equation}
C_l^i = \int dz\,
W^2_i(z) {H(z) \over d_A^2(z)} P_{hh}\left(\frac{l}{d_A}; z\right) \,,
\label{eqn:cl}
\end{equation}
where $W_i(z)$ is the distribution of halos in a 
given redshift bin normalized
so that 
$\int\, dz\, W_i(z)=1$,
$H(z)$ is the 
Hubble parameter, and 
$d_A$ is the
angular diameter distance in comoving coordinates.
Note that $W_i(z)$ comes directly from the observations of the number
counts as a function of redshift.

\begin{figure}[t]
\plotfiddle{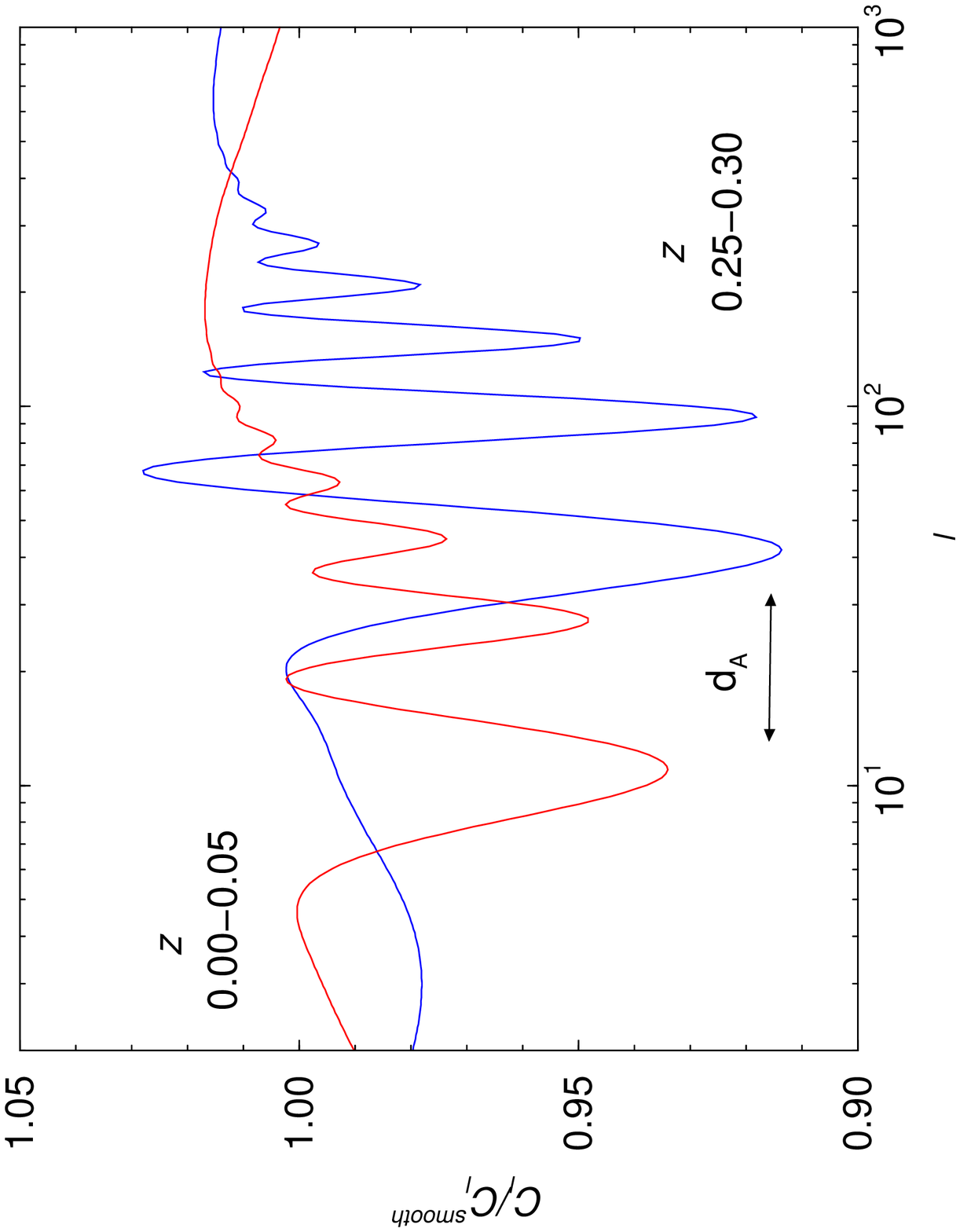}{1.0in}{-90}{33}{33}{-200}{100}
\plotfiddle{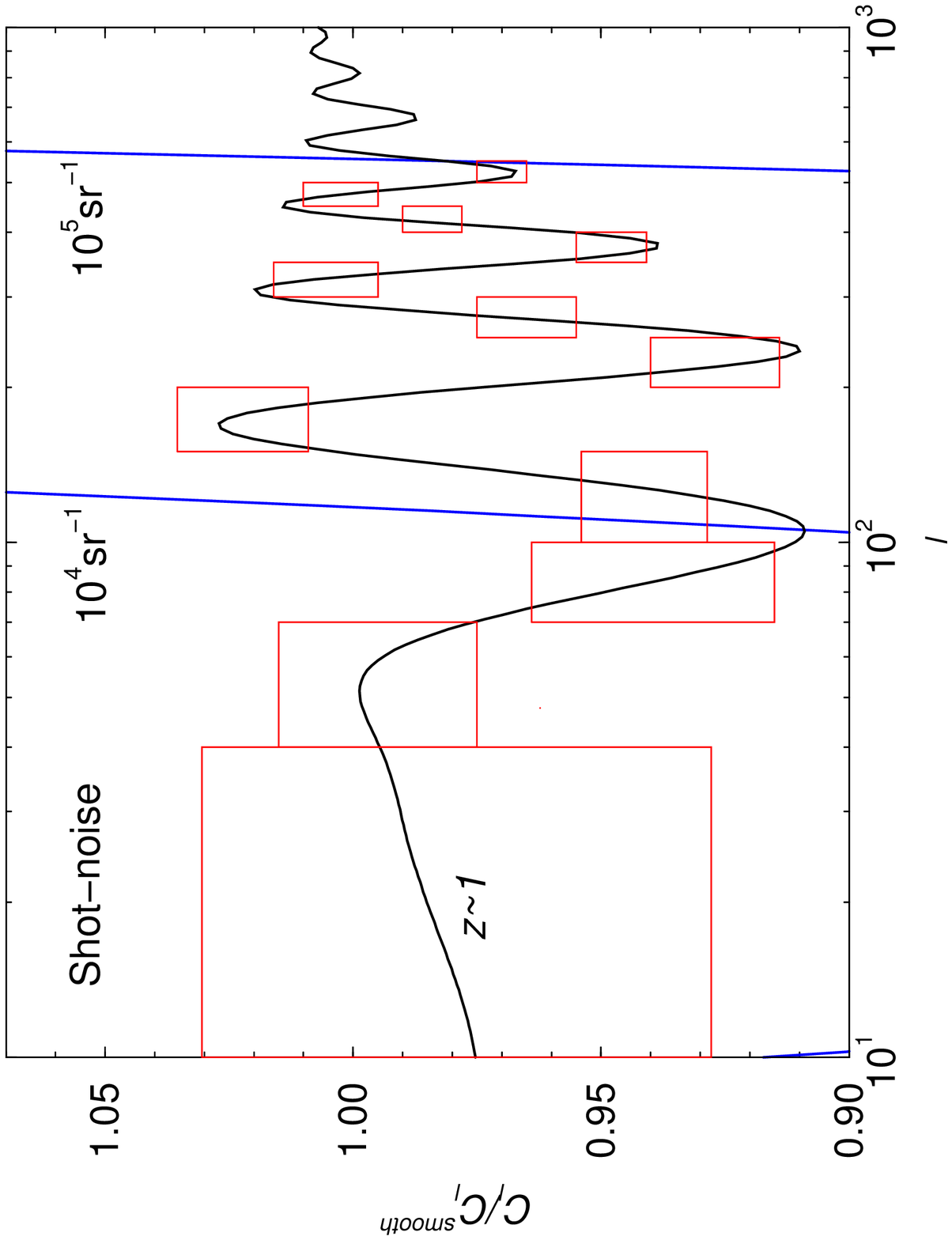}{1.0in}{-90}{33}{33}{-10}{185}
\caption{{\it Left:} The normalized power spectrum of 
halos at z $\sim$ 0.05 and z $\sim$ 0.3. The shift in oscillations is due to projection and correspond to change in distance between the two redshifts.
{\it Right:} Expected errors for the oscillations due to sample variance
alone. The shot-noise contribution due to the finite number of tracers are 
shown with a line. One requires $10^{5}$ sr$^{-1}$ source or more 
per z-bin for oscillation detection; this corresponds to a mass scale of
$\sim 10^{13}$ at $z \sim 1$.}
\end{figure}

The underlying linear power spectrum of density fluctuations, traced by halos,
 contains two physical scales:
the horizon at matter radiation equality $
k_{\rm eq} = \sqrt{2 \Omega_m H_0^2 (1+z_{\rm eq})} \propto \Omega_m h^2$,
which controls the overall shape of the power spectrum, and the sound 
horizon at the end of the Compton drag epoch, $k_{\rm s}(\Omega_m
h^2,\Omega_b h^2)$, which controls the small wiggles in the power spectrum.
The angular or multipole locations of these features shift in
redshift as $l_{\rm eq, s} = k_{\rm eq, s} d_A(z_i)$.
We propose the following test: measure
$C_l^i$ in several redshift bins and, using the fact that 
 $l_{\rm eq,s}$  scales with $d_A(z_i)$, 
constrain the  angular diameter distance as a function of redshift. 
Unlike the case with CMB, we can use tracers over a wide range in redshift
and measure the distance as a function of redshift, or redshift bin. 

The test can be affected by two ways:
(1) non-linearities generally erase the baryon oscillations and at $z < 1$ oscillations are generally affected. At low redshifts, one can use the test with the overall shape defined by $k_{\rm eq}$. (2) The tracer bias can be 
scale dependent, but for mass selected catalogs via lensing or SZ, 
bias can be studied with simulations.
\acknowledgments Author thanks Wayne Hu and collaborators, regrets lack of references, and thanks Fairchild foundation and the DOE for funding.

\end{document}